\documentclass[%
 reprint,
 amsmath,amssymb,
 aps,
]{revtex4-2}

\usepackage{graphicx}% Include figure files
\usepackage{dcolumn}% Align table columns on decimal point
\usepackage{bm}% bold math
\usepackage{hyperref}
\usepackage{amsmath}
\usepackage{graphicx}
\usepackage{xcolor}
\usepackage[separate-uncertainty=true,multi-part-units=single]{siunitx}

\usepackage{comment}

\begin{document}

\preprint{APS/123-QED}

\title{Neutron Stars as Perfect Fluids: Extracting the Linearized Response Function}% 
\author{Irvin Martínez-Rodríguez}
\affiliation{Department of Physics, Carnegie Mellon University, Pittsburgh, PA 15213, USA}
\email{ifmartin@andrew.cmu.edu }

\begin{abstract}
We develop an effective field theory framework for the conservative linear tidal response of a relativistic neutron star modeled as a perfect fluid with conserved particle number. Starting from the covariant fluid action in a curved background, we linearize about a static equilibrium configuration and obtain the quadratic action for fluid displacements coupled to metric perturbations. We then split the metric perturbation into induced and externally sourced parts, and integrate out the induced metric on a conservative matched domain using a symmetric Green kernel. The resulting effective fluid theory contains metric-mediated fluid interactions through a self-adjoint operator acting on the displacement, while the external perturbation acts as a tidal source. This self-adjoint structure permits a modal expansion of the pole sector: after projection onto conservative eigenmodes, the dynamics reduce to tidally driven oscillators with couplings fixed by relativistic inner products and overlap integrals. Matching these oscillator variables to the quadrupolar worldline theory gives analytic expressions for the modal dynamical tidal deformabilities in terms of mode frequencies, normalizations, and overlap integrals. We also identify a source-only non-pole sector generated by terms quadratic in the external perturbation. This sector does not affect the mode amplitudes, but its electric-quadrupolar projection can contribute to the full tidal response. This formulation clarifies which part of the relativistic tidal response is captured by explicit fluid modes and where a possible source-only non-pole contribution enters the worldline description.
\end{abstract}

%\keywords{Neutron Stars}%Use showkeys class option if keyword
                              %display desired
\maketitle

%\tableofcontents

\section{\label{sec:Intro}Introduction}

\begin{figure*}
\includegraphics[width=\textwidth]{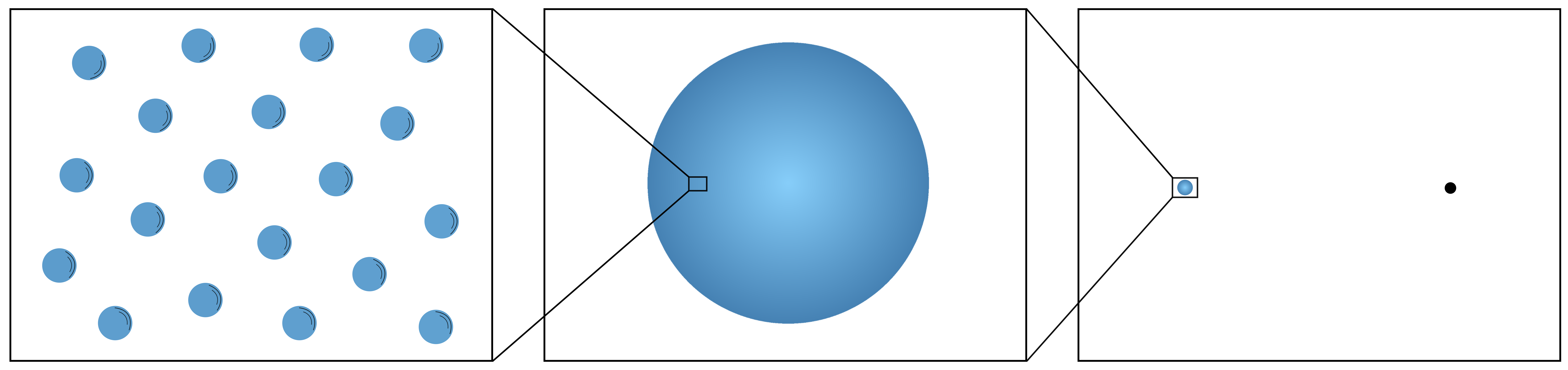}
\caption{\label{fig:wide}Schematic illustration of the hierarchy of scales involved in the problem. On the right-hand panel there is a well separated binary, where the point particle approximation is valid. In the middle panel we have zoomed in on the point-like particle, to find the star with radius $R_{\star}$. On the left panel, we have the hydrodynamical scale of the neutron perfect fluid.}
\end{figure*}

Extracting the response function of a neutron star is of significant scientific importance because it probes the star's internal dynamics \cite{Flanagan:2007ix}. However, formulating this problem covariantly presents theoretical challenges \cite{Poisson:2023xsr}. In this work, we show how effective field theory techniques provide a systematic way to address these challenges. We work with the conservative perfect-fluid effective theory \cite{Dubovsky:2005xd,Dubovsky:2011sj} in a curved background \cite{Delacretaz:2014oxa} and match it to a worldline effective theory \cite{Goldberger:2004jt,Goldberger:2005cd}, assuming the hierarchy of scales shown in Fig.~\ref{fig:wide}.

Within linearized general relativity, the adiabatic tidal response of a neutron star is encoded in its relativistic tidal deformabilities \cite{Hinderer:2007mb,Damour:2009vw,Binnington:2009bb}. The relativistic oscillation spectrum of nonrotating stars has been studied extensively in stellar perturbation theory \cite{Lindblom:1983ps,Detweiler:1985zz}, while explicit mode-sum formulations of the dynamical response were developed most systematically in Newtonian gravity \cite{Lai:1993di,Chakrabarti:2013xza,Pnigouras:2022zpx}. More recently, the frequency-dependent tidal response has been generalized systematically to general relativity using mode-less formulations \cite{Poisson:2023xsr}, near-zone matching \cite{Andersson:2025iyd}, matched relativistic perturbation theory \cite{HegadeKR:2024agt}, and worldline-EFT matching approaches, including scattering-amplitude matching \cite{Saketh:2026trm} and perturbative low-frequency matching \cite{Apostolidis:2026qsg}. Analytic-continuation methods have also been used to formulate a relativistic mode-expansion description with forced oscillator amplitudes and overlap integrals \cite{HegadeKR:2025qwj}. A subsequent Regge--Wheeler-gauge implementation of this mode-sum construction clarified both the robustness of the dominant \(f\)-mode response and the limitations of a purely modal representation \cite{HegadeKR:2026kku}. Recent reviews on tidal deformation can be found in \cite{Chakraborty:2026qru,Rodriguez:2026iot}.

In this work, we derive a covariant spectral representation of the pole part of the linear dynamical response directly from the bulk perfect-fluid effective action and match it to a worldline description with dynamical quadrupolar degrees of freedom. This yields analytic expressions for the modal frequency-dependent tidal deformabilities in terms of mode frequencies, normalizations, and overlap integrals, together with a direct map between the bulk fluid variables and the worldline oscillator variables. The same matching also identifies a source-only sector, independent of the explicit quadrupoles, whose tidal projection parameterizes possible non-pole contributions to the full response.

The fluid degrees of freedom are the comoving coordinates $\phi^A$, which label the fluid elements in an internal space. In curved spacetime, the equilibrium fluid-label triad $E_i^{A}$ and the spatial components of the vierbein $e_i^{I}$ provide the maps between internal/comoving indices, coordinate indices, and local orthonormal spatial indices. These structures allow the construction of invariant operators for the perturbations. Expanding the covariant perfect-fluid action about a static equilibrium configuration, we obtain the quadratic action governing linear fluid displacements coupled to metric perturbations.

To treat the metric perturbation, we follow the matched conservative framework of Ref.~\cite{HegadeKR:2025qwj}. We split the metric perturbation into a component induced by the fluid displacement and an externally sourced component, with the latter defined by analytic continuation of the exterior tidal field into the stellar interior. On the conservative matched domain, we use a symmetric Green kernel for the induced metric perturbation and integrate out this induced piece using diagrammatic tools. The resulting representation keeps the fluid displacement as the explicit interior dynamical variable, while the gravitational sector is encoded implicitly through the induced-metric Green kernel and its couplings to the external metric perturbation.

In the conservative sector, we introduce a relativistic inner product under which the effective linearized fluid operator \(O_{IJ}\) is self-adjoint. Expanding in the corresponding conservative fluid-mode basis, the spatial dependence of the fluid perturbation can be integrated out. The external perturbation defines the driving force, whose projection onto this basis determines the overlap integrals. This yields an effective linear response problem for the fluid displacement amplitudes with a discrete spectrum of driven modes. Matching the resulting driven system to a quadrupolar worldline action gives the pole part of the general relativistic dynamical response. This construction is related to earlier Newtonian mode-based response formalisms~\cite{Chakrabarti:2013xza,Zhou:2025lzg}. The full matched response can also receive a source-only non-pole contribution, independent of the explicit mode amplitudes, which must be fixed by matching. This separation between the modal pole sector and the source-only non-pole sector makes explicit which part of the response is computed by the fluid-mode expansion and which part must be fixed by matching.

The approximations, limitations, and breakdown of the model are discussed in the appendix. A numerical implementation in the Regge-Wheeler gauge for the frequencies and tidal deformabilities will be presented elsewhere. This framework provides a systematic bridge between relativistic stellar perturbations and the tidal response parameters relevant for gravitational-wave modeling.  

Conventions: We use metric signature \((-+++)\) and units with \(c=1\), keeping Newton's constant \(G\) explicit. Spacetime coordinate indices are denoted by \(\mu,\nu,\ldots\), with coordinate time component \(0\), and spatial coordinate indices by \(i,j,k,\ldots\). Local orthonormal frame indices are denoted by \(a,b,c,\ldots\), with local time component \(\hat{0}\), and local spatial frame indices by \(I,J,K,\ldots\). Material, or comoving internal, spatial indices are denoted by \(A,B,C,\ldots\).

~

\section{\label{sec:SHOmodel} The worldline effective theory}

In the worldline effective field theory for extended objects, the compact star is represented by a point particle moving on a worldline $y^\mu(\lambda)$ with multipole moments \cite{Goldberger:2004jt}.
The dynamical quadrupoles $Q_{IJ}(\lambda)$ are defined on the worldline, with the local spatial indices $I,J$ being the components in a local orthonormal spatial triad $e_{\mu}^{I}(\lambda)$ comoving with the object \cite{Goldberger:2005cd}.
The external tidal field $E_{IJ}(\lambda)$ denotes the electric-type tidal tensor evaluated on the worldline in the same local frame. This choice fixes a body frame, so the action below is not written in a manifestly local-rotation-invariant form. We parameterize the worldline by the coordinate time $t$, so the effective action reads 
\begin{equation}
    S = \int d t  \Bigg(- M_{\star}  + L_Q [Q_{},\dot Q_{}] - \frac{1}{2} Q_{IJ} E^{IJ} \Bigg) + S_{\rm src}^{(2)} [E^{}]  ~.
    \label{eq:actionworldine}
\end{equation}
The first term describes the unperturbed point-particle mass. The $Q$-dependent terms describe the explicit modal, or pole, part of the quadrupolar response. The source-only functional $S_{\rm src}^{(2)}[E]$ is independent of the dynamical quadrupole and parameterizes possible non-pole contributions built directly from the external tidal field.

We solve the equations of motion formally for the $Q$'s using linear response theory. In a normal-mode representation, 
\begin{flalign}
\begin{split}
    Q_{IJ}  (\omega) =&  -\lambda_{\rm pole }(\omega) E_{IJ}(\omega)
\end{split}\label{eq:response}
\end{flalign}
\noindent with $\lambda_{\rm pole}(\omega)$ receiving contributions from the tower of mode frequencies $\omega_n$ and deformability parameters $\lambda_n$,  
\begin{flalign}
    \lambda_{\rm pole} (\omega) = \sum_n \frac{\omega^2_{n} \, \lambda_n}{\omega^2_{n} -(\omega + i \epsilon)^2 }
\end{flalign}
The conservative worldline Lagrangian for the $Q$'s is \cite{Flanagan:2007ix}, 
\begin{flalign}
L_{Q} = \sum_n \frac{1}{ 4\lambda_{n} \omega_{n}^{2}}\left( \dot{Q}^{IJ}_{n} \dot{Q}^{IJ}_{n}-\omega_{n}^{2}\, Q^{IJ}_{n} Q^{IJ}_{n}\right) ~.
\end{flalign}
The source-only contribution in frequency space is
\begin{flalign}
S_{\rm src}^{(2)}[E]
=
\int \frac{d\omega}{2\pi}\,
\frac{1}{4}
\lambda_{\rm src}(\omega)
E_{IJ}(-\omega)E^{IJ}(\omega).
\end{flalign}
This term does not modify the equations of motion for the dynamical quadrupoles. It contributes instead to the on-shell source functional and therefore to the response extracted by matching. The coefficient $\lambda_{\rm src}(\omega)$ parameterizes a possible non-pole contribution to the full response. After the explicit quadrupoles are put on shell, the full electric-quadrupolar response is organized as
\begin{flalign}
    \lambda_{\rm full}(\omega)
    =
    \lambda_{\rm pole}(\omega)
    +
    \lambda_{\rm src}(\omega).
\end{flalign}

The worldline theory is organized as a low-frequency derivative expansion in \(\omega\), corresponding to an expansion of the response about \(\omega=0\). Introducing explicit oscillator variables resums the pole part of this expansion by retaining the resonant structure of the modes included in \(\lambda_{\rm pole}(\omega)\), thereby capturing their dynamics near resonance within linear response. In practice one truncates the mode sum to a finite set; the accuracy at higher frequencies is then limited by the lowest-frequency mode not included in the truncation and deteriorates as \(\omega\) approaches the next neglected resonance. The source-sector coefficient \(\lambda_{\rm src}(\omega)\) parameterizes a possible non-pole contribution and must be fixed by matching. Nonlinearities in the stellar response provide an additional, independent limitation when they become important.

\section{\label{sec:PFcurved}Perfect fluids in curved backgrounds}

We consider a relativistic compact object in static equilibrium, with a perfect-fluid interior and a spherically symmetric background. In a $3+1$ decomposition with vanishing shift, the interior metric takes the form
\begin{flalign}
    ds^2 = g_{\mu \nu} dx^{\mu} dx^{\nu} = - \alpha^2 dt^2 + \gamma_{ij} dx^{i} dx^{j},
    \label{eq:sphericalmetricGR}
\end{flalign}
\noindent where $\alpha  = \sqrt{-g_{00}}$ is the background lapse function and $\gamma_{ij}$ is the background spatial metric. For the metric perturbation, we write
\begin{flalign}
    h_{\mu\nu}=h_{\mu\nu}^{\rm in}+h_{\mu\nu}^{\rm ext},
    \label{eq:metricsplit}
\end{flalign}
where the induced piece $h_{\mu\nu}^{\rm in}$ is part of the dynamical response, while the external piece $h_{\mu\nu}^{\rm ext}$ is treated as a prescribed tidal source. This split is not unique inside the star. In the matched construction of Ref.~\cite{HegadeKR:2025qwj}, this ambiguity is fixed by identifying $h_{\mu\nu}^{\rm ext}$ with the interior continuation of the exterior post-Newtonian tidal field. This continuation is chosen to satisfy the tidal Hamiltonian and momentum constraints and to match the exterior tidal solution in the buffer region. With this prescription, $h_{\mu\nu}^{\rm ext}$ sources the dynamical response variables $h_{\mu\nu}^{\rm in}$ and the fluid displacement.

\subsection{Fluid degrees of freedom}
The low-energy dynamics of a perfect fluid are governed by symmetry principles \cite{Dubovsky:2005xd, Nicolis:2013lma}.
For a static equilibrium configuration we choose the comoving labels to align with the spatial coordinates in the internal space, $\phi^A = z^A (\boldsymbol{x})$. The relevant internal symmetries for a perfect fluid with conserved particle number are \cite{Dubovsky:2011sj} 
\begin{flalign}
    \phi^A &  \rightarrow \; \phi^A  + a^A,\\
    \phi^A &\rightarrow \; \mathcal{R}^A_{\;B} \phi^B,      \\
    \phi^A &\rightarrow \; \zeta^A (\phi),  \quad \textrm{with}  \quad \det(\partial \zeta^A / \partial \phi^B) = 1, \\
    \psi \; &\rightarrow \; \psi + f(\phi^I).
\end{flalign}
\noindent The first line is invariance under constant shifts. The second line is invariance under internal rotations, with $\mathcal{R}^A_{\;B}$ an $SO(3)$ matrix. The third line is invariance under volume-preserving diffeomorphisms, encoding the ability of the fluid to undergo adiabatic deformations. The last line is the comoving $U(1)$ symmetry for the conserved charge, with phase $\psi$.

Spacetime diffeomorphism invariance together with internal volume preserving diffeomorphisms implies that the operator $B^{AC} = g^{\mu \nu} \partial_{\mu} \phi^A \partial_{\nu} \phi^C$ enters the Lagrangian only through its determinant. Therefore, symmetry considerations restrict the covariant effective action of a perfect fluid to the form \cite{Dubovsky:2011sj} 
\begin{flalign}
S = \int d^4 x \, \sqrt{- g} \; F(b,y) 
\label{eq:fluidaction}
\end{flalign}
\noindent with $F(b,y)$ a function of $b = \sqrt{\det B^{AC}}$ and $y=u^{\mu} \nabla_{\mu} \psi$, which corresponds to a particular equation of state.

The building block $b$ is related to the entropy (comoving-volume) current $J^{\mu}$
and the four-velocity $u^{\mu}$.
Here $J^\mu$ is the identically conserved current constructed from the comoving scalars $\phi^A$ \cite{Dubovsky:2011sj}, hence
\begin{flalign}
J^{\mu} = b\,u^{\mu}, \qquad \quad  b = \sqrt{-g_{\mu \nu}J^{\mu} J^{\nu}} \, .
\end{flalign}
\noindent so that $u^{\mu}u_{\mu}=-1$ and $\nabla_{\mu} J^{\mu}=0$.  The conserved $U(1)$ charge current is comoving with the fluid current,
\begin{flalign}
j^\mu = F_y u^\mu = \frac{F_y}{b} J^{\mu},
\end{flalign}
so that $q \equiv F_y / b$ is advected with the fluid flow. 

A useful thermodynamic dictionary follows by comparing the stress-energy tensor derived from
eq.~(\ref{eq:fluidaction}) with the perfect-fluid form
\begin{flalign}
T_{\mu \nu} = (\epsilon+p)u_{\mu} u_{\nu} + p g_{\mu \nu}.
\end{flalign}
One finds the entropy density $s=b$ and temperature $T=-F_b$ (with $F_b = dF/db$), while the conserved particle-number density and chemical potential are
$n=F_y$ and $\mu=y$  \cite{Dubovsky:2011sj}. The rest-frame energy density and pressure are
\begin{flalign}
\epsilon = y F_y -F,
\qquad
p = F - bF_b,
\end{flalign}
so the enthalpy density is
\begin{flalign}
w\equiv \epsilon+p = y F_y  - bF_b.
\end{flalign}
Equivalently, $\epsilon+p = Ts +\mu n$.

In the presence of an external perturbation, the fluid is displaced according to
\begin{flalign}
\phi^A (x) = z^A (\boldsymbol{x})+ \pi^A (x),
\end{flalign}
and the functional $F(b,y)$ is expanded around the equilibrium configuration characterized by the background values $b_0 = \operatorname{det}(\partial_k z^{A}) \sqrt{\operatorname{det}g^{ij}}$ and $y_0$. The Lagrangian perturbation for $b$ to linear order is given by 
\begin{flalign}
    \Delta b = b_0 (\partial z^{-1})^{i}_{\,I}\,\partial_i \pi^I.
\end{flalign}
Since the particle-number current is comoving with the entropy current, the particle number per entropy density satisfies $\Delta q =0$, which to linear order implies the relation
\begin{flalign}
F_{yy}\Delta y
=
\left(
\frac{F_y}{b_0}-F_{by}
\right)\Delta b .
\end{flalign}
Therefore the perturbation of the chemical-potential variable $y$ is not an independent degree of freedom, but is fixed by the compressional perturbation $\Delta b$ through the comoving particle-number condition. 

Defining the equilibrium fluid-label triad $E_{i}^{\;A} \equiv \partial_{i } z^A$, the Eulerian perturbation of the fluid energy density follows from $\delta \epsilon = \delta (F_y y - F)$, with $\delta b$ determined using $\nabla_\mu J^\mu=0$. We denote the fluid-displacement contribution to this perturbation by $\delta\rho$. Then
\begin{flalign}
\begin{split}    
    \delta \rho_{} 
    =& \;  \nabla_i ( w_0 E^{i}_{A} \pi^A) - E^{i}_{A} \pi^A  \partial_i p_0 , 
    \label{eq:fluidpert}
\end{split}
\end{flalign}
with $w_0 = \epsilon_0 + p_0$. The full Eulerian energy-density perturbation is therefore $\delta \epsilon = \delta \rho - \tfrac{1}{2} w_0 g^{ij} h_{ij}$.

We formulate the action in the Eulerian frame. Expanding around the equilibrium background, the quadratic effective action for the fluid displacement $\pi^A$, including the linear coupling to metric perturbations, reads 
\begin{widetext}
\begin{flalign}    
\begin{split}    
S_{} = \int dt d^3 x \sqrt{-g}\Big(&  - \frac{w_0}{2}  \; g^{00} (B_0^{-1})_{AC} \dot{\pi}^{A}    \dot{\pi}^{C} - \frac{c_s^2}{2 w_0}   \, ( \delta \rho )^2 + \frac{1}{2 } \delta \rho  \left( -   g^{00} h_{00} + c_s^2 \, g^{ij } h_{ij} \right)  - w_{0}\,  g^{00} \dot{h}_{j0} E^j_A \pi^{A}    \Big) + \dots 
\label{eq:lagrangianframeactionperturbationInnermetricv3r}
\end{split}
\end{flalign}
\end{widetext}
\noindent  where overdots denote derivatives with respect to time $t$. We have used the thermodynamic variables and identified the speed of sound $c_s^2 = (-b^2 F_{bb} + \frac{(F_y - b F_{by})^2}{F_{yy}})/w_0$. We also define $B_{0}^{AC} \equiv \delta^{IJ} e^{i}_{I} e^{j}_{J}  E_{i}^{\;A} E_{j }^{\; C}$, where the spatial components of the background vierbein satisfy $g_{i j} = \delta_{IJ} e_{i}^I e_j^{ J}$. The last term has been rewritten using integration by parts in time, dropping the endpoint term under the usual fixed-endpoint variational boundary conditions. The ellipsis denotes metric-metric perturbations coming from the expansion of eq.~(\ref{eq:fluidaction}), whose explicit form is not needed for our discussion. To make the connection to the usual fluid displacement $\xi$ in the stellar perturbation literature, we identify $\xi^A = - \pi^A$.

To carry out the matching procedure to the worldline theory, we perform the field redefinition, $\pi^A \rightarrow - \alpha  E_{j}^{A} e^j_I \xi^I$. In the Eulerian frame, the fluid displacement equation of motion becomes
\begin{flalign}
\begin{split}
    w_0  \ddot{\xi}_I  
    =&  w_0  \alpha e^{i}_{I} \Bigg(   -\tilde{\nabla}_i  \left( \frac{c_s^2}{w_0} \delta \epsilon + \frac{1}{2} g^{00} h_{00}  \right)    +  g_{ki} \dot{h}^{k0} \Bigg)
    \label{eq:EOMlagr}
\end{split}
\end{flalign}
\noindent where $\tilde{\nabla}_i V \equiv \partial_i V + V \partial_i p_0 / w_0 $. Denoting
\begin{flalign}
    \delta U = \frac{c_s^2}{w_0} \delta \epsilon + \frac{1}{2} g^{00}h_{00},
\end{flalign}
\noindent the static tidal limit of eq.~(\ref{eq:EOMlagr}) gives \(\delta U=0\), reproducing the expected relation between the metric and energy-density perturbations \cite{Lindblom:1997un,Damour:2009vw}. Together with the metric perturbation equations, eq. (\ref{eq:EOMlagr}) determines the coupled eigenvalue problem. Solving this system gives the mode frequencies and eigenfunctions, which determine the inner product and overlap integrals used in the matching.

\subsection{Gravitational degrees of freedom}
\label{subsection:gravityDOF}

Inside the star, we begin from the full Einstein--Euler action
\begin{flalign}
\begin{split}
S=\int d^4x \sqrt{-g}\left(\frac{R}{16\pi G}-\epsilon\right),
\end{split}
\end{flalign}
whose variation yields the nonlinear Einstein equations together with the fluid equations. Linearizing about a static background star, the metric perturbations satisfy
\begin{flalign}
\delta G_{\mu\nu} =8\pi G\,\delta T_{\mu\nu},
\label{eq:linearEinsteinEuler}
\end{flalign}
with $\delta T_{00} =  \left( \alpha^2 \delta \epsilon - \epsilon_0   h_{00} \right)$, $\delta T_{0i} = - w_0 g_{ij} \alpha e^{j}_J \dot{\xi}^J - \epsilon_0  h_{0i}$ and $\delta T_{ij} = g_{ij} c_s^2 \delta \epsilon + p_0 h_{ij}$. In the canonical formulation, eq.~(\ref{eq:linearEinsteinEuler}) together with the perturbed Euler equation, i.e. the orthogonal projection of $\delta(\nabla_\mu T^{\mu\nu})=0$, defines the linearized Einstein--Euler system. In our formulation, the fluid dynamics is represented by the displacement equation of motion, eq.~(\ref{eq:EOMlagr}). Therefore, eq.~(\ref{eq:EOMlagr}) together with eq.~(\ref{eq:linearEinsteinEuler}) determines the dynamics of the coupled system.

Using the decomposition in Eq.~(\ref{eq:metricsplit}) and the analytic continuation of the external tidal field into the stellar interior, Ref.~\cite{HegadeKR:2025qwj} analyzes the matched conservative Einstein--Euler problem for the dynamical variables $(\xi^{\mu},h_{\mu\nu}^{\rm in})$, with the external tidal perturbation treated as a source. By matching the strong-field stellar solution to a post-Newtonian exterior solution in a buffer region, and working in harmonic gauge, the linearized Einstein--Euler system can be written in terms of a self-adjoint operator acting on the dynamical perturbations. The Hamiltonian and momentum constraints restrict the perturbations entering the matched Einstein--Euler problem, while the buffer-zone matching supplies the conservative tidal boundary conditions. A subsequent analysis showed that, in Regge--Wheeler gauge, the dominant \(f\)-mode contribution provides a robust approximation to the direct matching result, even though a purely modal representation of the full relativistic response is subtle: the inner-product operator is not positive definite on the full Regge--Wheeler-gauge function space, so the corresponding relativistic mode sum is not expected to converge strictly to the full matched response \cite{HegadeKR:2026kku}.

To avoid relying on a complete mode expansion of the induced metric perturbation, in the present work we use the same conservative, gauge-fixed matched domain and assume that the induced metric sector admits a symmetric Green kernel,
\begin{flalign}
G_{\mu\nu\rho\sigma}^{h^{\rm in}h^{\rm in}}(x,x')
=
G_{\rho\sigma\mu\nu}^{h^{\rm in}h^{\rm in}}(x',x).
\end{flalign}
This Green kernel is understood as the inverse of the induced-metric block of the gauge-fixed conservative quadratic operator on the constrained matched domain, with the constraints and matched boundary conditions imposed. It allows us to integrate out \(h_{\mu\nu}^{\rm in}\) and obtain a dressed effective operator for the fluid displacement. In block form, this gives the schematic effective action 
\begin{flalign}
    S_{\rm eff}[\xi,h^{\rm ext}] = \frac{1}{2} \xi O_{\rm  eff} \xi +\xi J_{} [h^{\rm ext}] +\frac{1}{2} h^{\rm ext} D_{\rm eff} h^{\rm ext}
\end{flalign}
where
\begin{flalign}
O_{\rm eff}
=&
\,O_{\xi\xi}
-
O_{\xi h^{\rm in}}
\,G_{h^{\rm in}}\,
O_{h^{\rm in}\xi}\\
D_{\rm eff}=& \, O_{h^{\rm ext} h^{\rm ext}}-O_{h^{\rm ext} h^{\rm in}}\,G_{h^{\rm in}}\,O_{h^{\rm in}h^{\rm ext}},
\end{flalign}
with $G_{h^{\rm in}} =(O_{h^{\rm in}h^{\rm in}})^{-1}$. Thus the induced gravitational sector is retained implicitly through the induced-metric Green kernel, which dresses the fluid operator and the couplings to the external tidal field. In this way, the explicit induced metric variables are eliminated, while the matched conservative boundary conditions remain encoded in the symmetric Green kernel. Consistent with Ref.~\cite{HegadeKR:2026kku}, we do not require a convergent mode expansion of the induced metric perturbation itself; the explicit spectral decomposition is applied only to the effective fluid-displacement problem. With the response defined through the modal quadrupoles in Eq.~(\ref{eq:response}), the term \(h^{\rm ext}D_{\rm eff}h^{\rm ext}\) is independent of \(\xi\) and therefore does not affect the modal amplitude equations. It can, however, contribute to the full matched source response as a source-only non-pole term.

For the external metric perturbation, we keep $h_{\mu\nu}^{\rm ext}$ as a prescribed source and define the tidal field directly from it. After matching to the worldline theory, the hierarchy of scales between the stellar radius and the binary separation allows us to use the framework of non-relativistic general relativity~\cite{Goldberger:2004jt}. In the conservative sector, the external tidal field is generated by potential graviton exchange with the companion in the weak-field limit. The corresponding propagators are discussed in Appendix~\ref{app:metricpert}.

~

\subsection{Effective gapped fluid dynamics}

After the field redefinition
\begin{flalign}
    \pi^A \rightarrow - \alpha E_j^A e^j_I \xi^I,
\end{flalign}
and after integrating out the induced metric perturbation using diagrammatic tools, the effective quadratic Lagrangian density for the fluid displacement in the Eulerian frame is written as
\begin{flalign} 
\begin{split}    
\mathcal{L}_{\xi} =&
\frac{w_0}{2}
\left(
\delta_{IJ}\dot{\xi}^I\dot{\xi}^{J}
-
\xi^I O_{IJ}\xi^J
\right)
+\frac{1}{2} \delta T_{\xi}^{\mu\nu}h_{\mu\nu}^{\rm ext},
\end{split}
\end{flalign}
where \(\delta T_\xi^{\mu\nu}\) denotes the part of the stress-tensor perturbation linear in the fluid displacement, with
\begin{flalign}
\delta T_{\xi}^{\mu\nu}h_{\mu\nu}^{\rm ext} = \delta\rho \left(\alpha^{-2}h_{00}^{\rm ext}
+ c_s^2g^{ij}h_{ij}^{\rm ext}\right)
+ 2 w_0 \alpha^{-1} h_{0i}^{\rm ext} e^i_I \dot{\xi}^I.
\end{flalign}
In defining the operator \(O_{IJ}\), we have integrated by parts at the level of the action, with boundary terms vanishing under the stellar surface conditions. The fluid perturbation is
\begin{flalign}
\begin{split}
\delta\rho
= -\nabla_i\left(w_0\alpha e^i_I\xi^I\right)
+ \alpha e^i_I\xi^I\partial_i p_0.
\end{split}
\end{flalign}

The quadratic form defines a gravity-dressed self-adjoint operator acting on \(\xi^I\),
\begin{widetext}
\begin{flalign}
O_{IJ}  \xi^J (x) = \alpha e^i_I\Bigg(    \nabla_i \left( \frac{c_s^2}{ w_0 }  \delta \rho \right)  + \frac{c_s^2 \partial_i p_0}{w_0^2} \delta \rho +  \frac{1}{2} \nabla_i \left(\int dt' d^3x' \sqrt{-g (x')} g^{00}(x) \langle h_{00}^{\rm in}(x) h_{00}^{\rm in}(x')\rangle  g^{00}(x')  \delta \rho (x') \right) \Bigg) + \dots ~.
\label{eq:operatorOxi}
\end{flalign}
\end{widetext}
This operator is self-adjoint with respect to the inner product weighted by \(\sqrt{-g}w_0\), and \(\langle\cdot\rangle\) denotes the classical conservative Green kernel. The remaining terms in the ellipsis encode the other metric-mediated contributions from integrating out \(h_{\mu\nu}^{\rm in}\). They are built from the same symmetric conservative Green kernel and are therefore self-adjoint; their explicit form is shown in Appendix~\ref{app:metricpert}.

Assuming completeness of the conservative fluid-mode basis, the fluid displacement can be expanded in terms of time-dependent mode amplitudes \(q_n(t)\) and spatial eigenfunctions \(\xi_n^I=\xi_n^I(\boldsymbol{x})\) of the self-adjoint operator \(O_{IJ}\):
\begin{flalign}
    \xi^I(t,\boldsymbol{x})
    =
    \sum_n q_n(t)\,\xi_n^I(\boldsymbol{x}) .
\end{flalign}
The normal modes satisfy
\begin{flalign}
    O_{IJ}\xi_n^J
    =
    \omega_n^2\,\delta_{IJ}\xi_n^J,
\end{flalign}
with real eigenvalues \(\omega_n^2\). The eigenfunctions are orthogonal with respect to the weighted inner product
\begin{flalign}
    \boldsymbol{\xi}_m  \cdot \boldsymbol{\xi}_n 
    =
    \int d^3x\,\sqrt{-g}\,w_0\,
    \delta_{IJ}\xi_m^I \xi_n^J = N_{n} \delta_{m n},
\end{flalign}
with normalization $ N_n = M_\star R_\star^2 \mathcal{N}_n $, where 
$\mathcal{N}_n$ is dimensionless.

Therefore, after projecting onto the normal-mode basis and integrating over the stellar volume, the effective action takes the form \(S=\int dt\,L\), with
\begin{flalign} 
\begin{split}    
L
=
\sum_n
\left(
\frac{N_n}{2}
\left(
\dot{q}_n^2-\omega_n^2q_n^2
\right)
+
\frac{1}{2}
\int d^3x\,\sqrt{-g}\,
\delta T_{\xi_n}^{ab}h_{ab}^{\rm ext}
\right).
\label{eq:fluidactiongapped}
\end{split}
\end{flalign}
The spatial integral defines the source functional for each mode. The local-frame components of the external metric perturbation are
\begin{flalign}
\begin{split}
    h_{\hat{0}\hat{0}}^{\rm ext} =&\, - g^{00}h_{00}^{\rm ext},\\
    h_{\hat{0} I}^{\rm ext} =&\, \alpha^{-1} e^i_I  h_{0i}^{\rm ext},\\
    h_{IJ}^{\rm ext} =& \, e_{I}^i e_J^{j} h_{ij}^{\rm ext}.
\end{split}
\end{flalign}
The stress-tensor perturbation generated by the \(n\)-th mode is given by 
\begin{flalign}
\begin{split}
    \delta T^{\hat{0}\hat{0}}_{\xi_n}
    =& \, q_n \delta\rho_n,\\
    \delta T^{\hat{0}I}_{\xi_n}
    =& \,
    w_0  \dot{q}_n\xi_n^I,\\
    \delta T^{IJ}_{\xi_n}
    =& \,
    c_s^2q_n\delta\rho_n \delta^{IJ}.
\end{split}
\end{flalign}

The source functional can then be rewritten in multipolar form by expanding \(h_{ab}^{\rm ext}\) around the center-of-mass worldline. This expansion identifies the induced quadrupole contribution \(Q^{IJ}\). In the local weak-field region, the leading electric-quadrupole coupling is matched to the worldline form \(-\frac{1}{2}Q^{IJ}E_{IJ}\).

\section{Matching the response function}
\label{sec:pfcurved}

We now match the mode-projected source in eq.~(\ref{eq:fluidactiongapped}) to the worldline tidal coupling. To isolate the quadrupolar response, we use the hierarchy between the stellar radius and the external-field variation scale and perform a local-frame multipole expansion of the external metric perturbation around the center-of-mass worldline,
\begin{flalign}
    h_{ab}^{\rm ext}(t,\boldsymbol{x})
    =
    \sum_{\ell=0}^{\infty}
    \frac{1}{\ell!}
    x^L
    \left(
    \nabla_L h_{ab}^{\rm ext}
    \right)(t,\boldsymbol{0}).
\end{flalign}
Here \(L=K_1\dots K_\ell\) is a spatial multi-index, \(x^L=x^{K_1}\cdots x^{K_\ell}\), with \(x^I\) local-frame spatial coordinates centered on the worldline, and \(\nabla_L=\nabla_{K_1}\cdots\nabla_{K_\ell}\) denotes the frame-covariant derivative acting on the local-frame tensor \(h_{ab}^{\rm ext}\), evaluated on the center-of-mass worldline.

The electric quadrupolar coupling receives contributions from the \(\ell=2\) term of \(h_{\hat0\hat0}^{\rm ext}\), the \(\ell=1\) term of \(h_{\hat0 I}^{\rm ext}\), and the \(\ell=0\) term of \(h_{IJ}^{\rm ext}\). This motivates defining the \(n\)-th mode induced quadrupolar moment  as 
\begin{flalign}
\begin{split}
    Q^{IJ}_n (t) =&  \int d^3 x  \sqrt{-g}\; \delta T_{\xi_n}^{\hat{0} \hat{0}}  \;  x^{\langle I}x^{J\rangle}\\ 
    =& \, q_n (t) \int d^3 x  \sqrt{-g}\;  \delta \rho_n (\boldsymbol{x})  \; x^{\langle I}x^{J\rangle}.
\end{split}
\end{flalign}
The curved stellar background enters this spatial overlap integral through the induced fluid density perturbation \(\delta\rho_n\) and the spatial measure.

For the purpose of rewriting the linear source coupling, we use the local conservation law for the unforced conservative mode,
\(\nabla_a \delta T_{\xi_n}^{ab}=0\). Using this relation, together with integrations by parts and the vanishing of the relevant spatial and temporal boundary terms, the electric quadrupolar part of the source coupling can be written as
\begin{flalign} 
\begin{split}    
\frac{1}{2}
\int dt  d^3x\,\sqrt{-g}\,
\delta T_{\xi_n}^{ab}h_{a b}^{\rm ext} = -\frac{1}{2} \int dt Q^{IJ}_n F_{IJ},
\end{split}
\end{flalign}
where 
\begin{flalign}
    F_{IJ} =  \frac{1}{2} \left(- \nabla_I \nabla_J h_{\hat{0} \hat{0}}^{\rm ext}  +   2 \nabla_{\hat{0}} \nabla_{(I} h_{J) \hat{0}}^{\rm ext}  -  \nabla_{\hat{0}} \nabla_{\hat{0}}  h_{I J}^{\rm ext}  \right).
\end{flalign}
Here \(F_{IJ}\) is the local-frame linearized tidal source combination constructed from the external metric perturbation. Its relation to the electric-type tidal tensor is made in the weak-field region below.

To evaluate the overlap integrals explicitly, we now specialize to a static, spherically symmetric perfect-fluid star. In Schwarzschild-like coordinates, the background metric can be written as
\begin{flalign}
     g_{\mu\nu}dx^\mu dx^\nu
     =
     -e^{2\Phi(r)}dt^2
     +
     e^{2\Lambda(r)}dr^2
     +
     r^2 d\Omega^2 ,
    \label{eq:trulysphericalmetricGR}
\end{flalign}
\noindent where \(e^{2\Lambda(r)}=(1-2Gm(r)/r)^{-1}\). The mass function \(m(r)\), the gravitational potential \(\Phi(r)\), and the background fluid profiles \(\epsilon_0(r)\) and \(p_0(r)\), and hence \(w_0(r)\), are determined by the TOV equations once an equation of state is specified~\cite{Oppenheimer:1939ne,Tolman:1939jz}.

On this static spherical background, the fluid displacement can be decomposed in a real vector-spherical-harmonic basis,
\begin{flalign}
    \xi^I(t,\boldsymbol{x})
    =
    \sum_{n\ell m}
    q_{n\ell m}(t)\,
    \xi_{n\ell m}^I(\boldsymbol{x}) .
\end{flalign}
The spatial eigenfunctions are expanded into radial (R), electric-type (E), and magnetic-type (B) vector spherical-harmonic components. Since the electric tidal field sources only the even-parity sector, only the radial and electric-type components contribute, and we set \(\xi_{n\ell}^{\rm B}=0\) for the modes considered below. The orthogonality relation in this basis is
\begin{flalign}
    \int d^3x\,\sqrt{-g}\,w_0\,
    \boldsymbol{\xi}_{n'\ell' m'}
    \cdot
    \boldsymbol{\xi}_{n\ell m}
    =
    N_{n\ell}\,
    \delta_{n n'}\delta_{\ell \ell'}\delta_{m m'} ,
    \label{eq:orthogonalityrel}
\end{flalign}
with normalization
\begin{flalign}
    N_{n\ell}
    =
    \int_0^{R_{\star}} dr\, e^{\Phi+\Lambda} w_0
    \left[
    \left(r \xi_{n\ell}^{\mathrm{R}}\right)^2
    +
    \ell(\ell+1)
    \left(r \xi_{n\ell}^{\rm E}\right)^2
    \right].
    \label{eq:normalization}
\end{flalign}

Substituting the mode expansion into the Lagrangian in Eq.~(\ref{eq:fluidactiongapped}), the fluid dynamics reduces to a collection of driven harmonic oscillators with eigenfrequencies \(\omega_{n\ell}\),
\begin{flalign}    
L  =&   \sum_{n \ell m}  \left( \frac{N_{ n \ell }}{2} \left( \dot{q}_{n \ell m}^2 - \omega_{n \ell }^2  \, q_{n \ell m}^2  \right) +  q_{n \ell m } f_{n \ell m}   \right) ~ .
\label{eq:hsoq}
\end{flalign}
Here $f_{n\ell m}$ is the driving force obtained by projecting the source functional onto the mode $(n,\ell,m)$. In the frequency domain, for the quadrupolar sector $\ell=2$, the retarded solution for the mode amplitude is
\begin{flalign}
\label{eq:fluid_amp_sol}
q_{n2m}(\omega)
=
\frac{1}{\omega_{n2}^2-(\omega+i\epsilon)^2}
\frac{f_{n2m}(\omega)}{N_{n2}} .
\end{flalign}

The induced quadrupole associated with a given mode can be written as
\begin{flalign}
    Q_{n2m}^{IJ}(t)
    =
    q_{n2m}(t)\,\mathcal Q_{n2m}^{IJ},
\end{flalign}
where \(\mathcal Q_{n2m}^{IJ}\) is the time-independent quadrupolar overlap tensor. In the real symmetric trace-free basis, the driving force is then
\begin{flalign}
\begin{split}
     f_{n2m}(t)
     &=
     -\frac{1}{2}\mathcal Q_{n2m}^{IJ}F_{IJ}(t)
     \\
     &=
     -\frac{1}{2}
     \sqrt{\frac{32\pi}{15}}\,
     I_{n2}\,
     \mathcal Y_{2m}^{IJ}F_{IJ}(t).
\end{split}
\end{flalign}
\noindent Here \(\mathcal{Y}_{\ell m}^{IJ}\) denotes a real spherical-harmonic symmetric trace-free tensor~\cite{Chakrabarti:2013xza}, and \(I_{n\ell}=M_\star R_\star^\ell \mathcal{I}_{n\ell}\) is the dimensionful overlap integral, with \(\mathcal{I}_{n\ell}\) dimensionless.

The fluid mode overlap integral is
\begin{flalign}
\label{eq:overlap_int}
I_{n \ell}=   \int_{0}^{R_{\star}} d r r^{\ell+1} e^{2\Phi}  w_0\Bigg( \Big( 1 -    \frac{r \Phi'}{\ell }   \Big) \xi_{n \ell}^{\mathrm{R}}+ (\ell+1) \xi_{n \ell}^{\mathrm{E}}\Bigg).
\end{flalign}
This expression is obtained after integrating by parts in the radial integral. The term proportional to \(\Phi'\) arises from the pressure-gradient contribution to the density perturbation, after using hydrostatic equilibrium. At this stage, the spatial fluid degrees of freedom have been integrated over the stellar radius, corresponding to the middle panel of Fig.~\ref{fig:wide}.

The driving field \(F_{IJ}\) is matched in the weak-field region, where the local-frame covariant derivatives reduce to ordinary derivatives at the center-of-mass worldline. In this limit, \(F_{IJ}\) is identified with the linearized electric-type tidal tensor,
\begin{flalign}
    E_{IJ}
    =
    \frac{1}{2}
    \left(
    - \partial_I \partial_J h_{00}^{\rm ext}
    + \partial_{0} \partial_{I} h_{J0}^{\rm ext}
    + \partial_{0}\partial_{J} h_{I0}^{\rm ext}
    - \partial_{0}^{2} h_{IJ}^{\rm ext}
    \right).
\end{flalign}
\noindent This is the linearized expression for $R_{I0J0}$, or equivalently the electric part of the Weyl tensor in the vacuum weak-field region, up to the normalization convention used for $h_{\mu\nu}^{\rm ext}$. In vacuum, $E_{IJ}$ is symmetric and trace-free. The external perturbation can then be generated by coupling the gravitational field to the companion worldline and integrating out potential graviton modes, as in non-relativistic general relativity~\cite{Goldberger:2004jt}. This diagrammatic matching is summarized in Appendix~\ref{app:metricpert}.

After inserting the solution for the time-dependent amplitude in Eq.~\eqref{eq:fluid_amp_sol},the modal contribution to the fluid response function can be extracted. Suppressing the quadrupolar label $\ell =2$, the modal tidal deformability $\lambda_{n}$ appearing in Eq.~(\ref{eq:response}), and the corresponding dimensionless Love number $k_n$, with dimensionless frequency $ \hat{\omega}_{n} = \sqrt{R^3_{\star}/G M} \omega_{n}$, are
\begin{flalign}
    \lambda_{n} = \frac{16 \pi}{15} \frac{1}{\omega^2_{n}}  \frac{I_{n }^2}{N_{n}} 
    \qquad \mathrm{and}  \qquad k_{n} = \frac{8 \pi}{5}  \frac{1}{\hat{\omega}_{n}^2} \frac{\mathcal{I}_{n }^2}{\mathcal{N}_{n}}. 
\end{flalign}
\noindent With these definitions, the effective action for the fluid amplitudes in Eq.~(\ref{eq:hsoq}) is matched to the worldline effective action in eq.~(\ref{eq:actionworldine}), completing the description of the perfect-fluid star as a worldline object with dynamical tidal degrees of freedom, as in the right panel of Fig.~\ref{fig:wide}. The amplitude solution in Eq.~\eqref{eq:fluid_amp_sol}, together with the eigenfunctions \(\boldsymbol{\xi}_{n\ell m}\), determines the linear stellar fluid response to the external tidal field.

\section{Discussion and Conclusions}
\label{sec:con}

In this paper we developed an effective field theory framework to extract the conservative sector of the general relativistic linear response of a perfect-fluid neutron star driven by externally sourced dynamical metric perturbations. The modal pole sector is naturally organized in terms of the star's normal-mode content: it is determined by the mode frequencies together with a set of overlap and normalization integrals that encode how each mode couples to the external tidal field. The full matched response can also contain a source-only non-pole contribution, which is independent of the explicit mode amplitudes.

Our formulation is complementary to recent relativistic treatments based on matched asymptotic expansions and analytic continuation of the linearized Einstein--Euler system, which describe the dynamical tidal response by matching strong-field stellar perturbations to a post-Newtonian tidal environment.

In the present formulation, the response is derived from the covariant bulk perfect-fluid effective action after integrating out the induced metric perturbation on a conservative matched domain. This viewpoint organizes the problem by scales and makes the matching to the worldline response function transparent. In practice, one first solves the coupled fluid--gravity eigenvalue problem to obtain the mode frequencies and eigenfunctions. The modal pole response is then extracted by projecting the effective source onto the fluid-displacement component of these modes, so that the modal dynamical tidal deformabilities are determined by the mode frequencies, normalization integrals, and overlap integrals. The remaining source-only sector is independent of the mode amplitudes and must be fixed by matching to the full source response. Quantitative results will be presented elsewhere. Extending the present conservative treatment to include dissipative effects will require a dedicated analysis.

The regime of validity is set by linear response and by the number of modes retained. In practice one truncates the mode sum; the accuracy at higher \(\omega\) is controlled by the lowest-frequency omitted mode and deteriorates as \(\omega\) approaches the corresponding neglected resonance. Separately, the model breaks down when nonlinearities in the stellar response become important, which can occur near resonance, \(\omega\sim\omega_n\), if the driven mode amplitude becomes sufficiently large. Appendix~\ref{app:approx} provides a parametric estimate of the leading nonlinear corrections by expanding the fluid perturbation action to \(O(\pi^3)\) and mapping the result onto a cubic \(Q^3\) interaction in the worldline effective action. Determining the precise breakdown point for close binaries would require evolving the driven mode amplitudes \(q_n(t)\), or equivalently the quadrupolar variables \(Q_n(t)\), along a binary inspiral and is beyond the scope of the present work.

More broadly, our approach provides a covariant description of neutron stars as relativistic perfect fluids coupled to external gravitational fields, applicable whenever a perfect-fluid description is adequate. In regimes where superfluid or multi-component effects are important, the bulk fluid effective theory should be replaced by the corresponding superfluid or multifluid theory. The same scale-separated matching logic should carry over, although the spectrum, overlap integrals, and worldline degrees of freedom may be modified. Extending the bulk theory to the superfluid case is currently under investigation.

\begin{acknowledgments}
IMR is particularly grateful to Ira Rothstein, Jinping Li and Abhishek Hegade. IMR also thanks Tejaswi Venumadhav, Riccardo Penco, Tanja Hinderer and Paolo Creminelli for insightful discussions, and acknowledges support from the US Department of Energy (HEP) Award (Grant No. DE-FG02-04ER41338 and DE-FG02-06ER41449)
\end{acknowledgments}

\appendix

\section{Integrating out metric perturbations}
\label{app:metricpert}

\begin{figure*}
\includegraphics[width= 1 \textwidth]{figures/fluiddiagramsv2.png}
\caption{Diagrammatic representation of the leading conservative metric-mediated interactions. The dashed vertical line denotes the fluid displacement, while the solid vertical line denotes the external point-mass source. In panel (a), the horizontal line denotes the induced-metric Green kernel mediating the interaction between two fluid perturbations inside the star. In panel (b), the horizontal line denotes the weak-field potential-mode propagator mediating the coupling between a fluid perturbation and the external source.
}
\label{fig:wide2}
\end{figure*}

\subsection*{Induced metric perturbation}

In the conservative matched problem discussed in the main text, the induced metric perturbation \(h_{\mu\nu}^{\rm in}\) is integrated out using the symmetric Green kernel associated with the linearized Einstein--Euler operator. The resulting effective theory is nonlocal in spacetime and encodes the gravitationally mediated interaction between fluid perturbations.

We denote the corresponding conservative kernel by 
$G_{hh}^{\rm sym}(x,x')\equiv G_{hh}^{\rm sym}(t,\boldsymbol{x};t',\boldsymbol{x}')$. In diagrammatic notation we write the induced metric propagator as
\begin{align}
    \begin{gathered}
     \includegraphics[width =2cm]{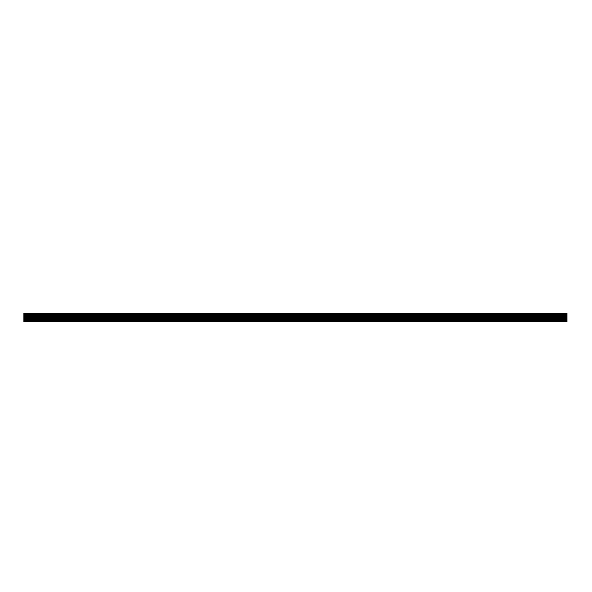}
    \end{gathered}
    = \langle h_{\mu \nu}^{\rm in}(x) h_{\rho \sigma}^{\rm in}(x')\rangle
    \equiv \, G_{\mu \nu \rho \sigma}^{h^{\rm in} h^{\rm in}, \rm sym}(x,x').
    \label{eq:propinside}
\end{align}
\noindent The induced-metric Green kernel is symmetric, $G^{h^{\rm in} h^{\rm in}, \rm sym}_{\mu \nu \rho \sigma}(x,x')=G^{h^{\rm in} h^{\rm in}, \rm sym}_{\rho \sigma \mu \nu}(x',x)$, which is the property needed for the induced contribution to define a self-adjoint effective fluid operator. 

The linear fluid--metric vertex entering the induced-metric exchange is
\begin{align}
    \begin{split}
        \begin{gathered}
   \includegraphics[width = 1.5cm]{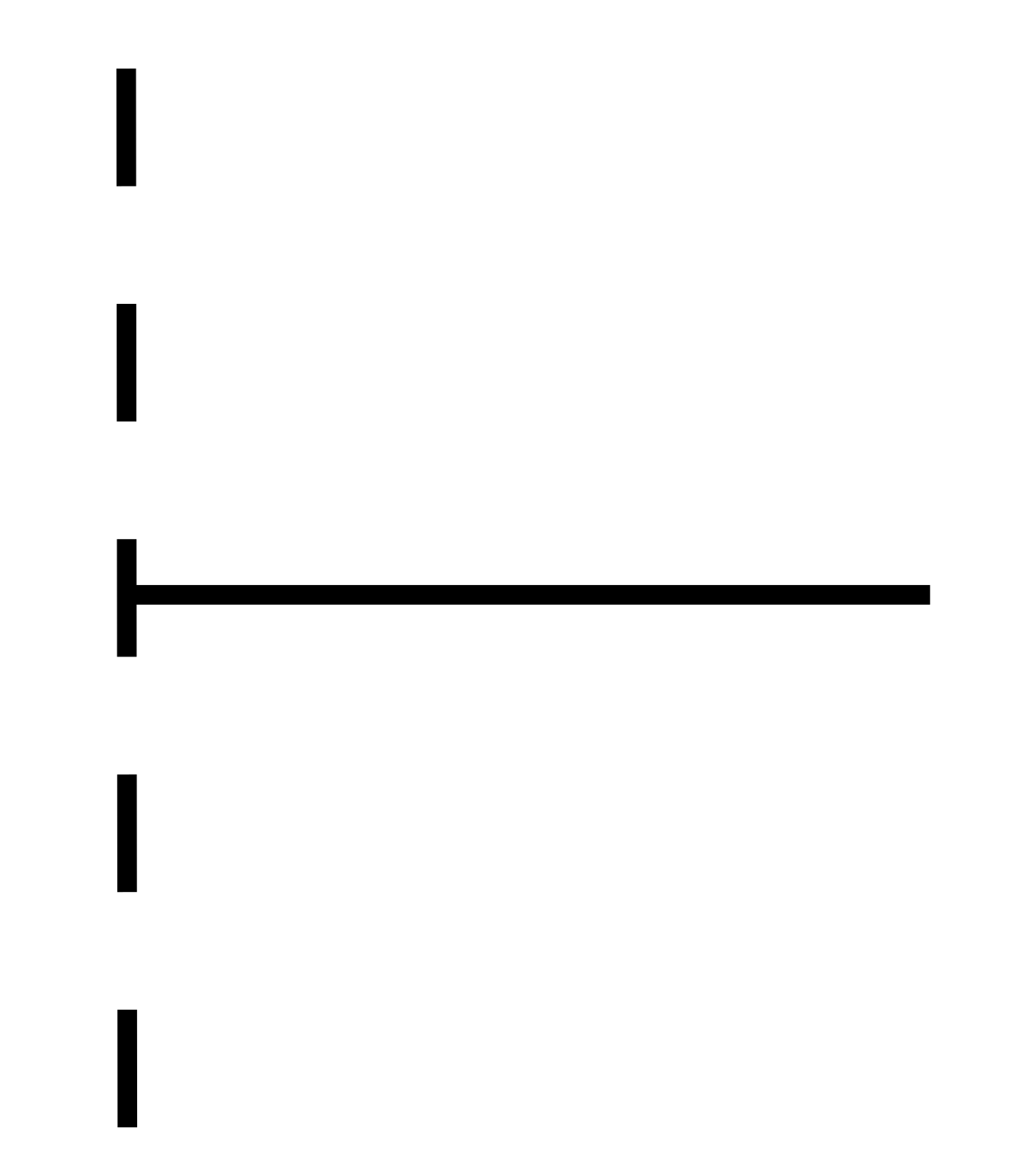}
        \end{gathered}  
        =&   \int d t d^3 x \, \sqrt{-g} \,\Big( -\frac{1}{2} g^{00} h_{00} \, \delta \rho \\
        & \quad \quad + \frac{1}{2}  c_s^2  \delta \rho \, g^{ij } h_{ij}  - w_{0} g^{00} \dot{h}_{i0}  \alpha e^{i}_I \xi^{I} \Big) 
    \end{split}
\label{eq:fluiddiagrams}
\end{align}    
\noindent In this case, the dashed vertical line represents the fluid displacement, while the horizontal solid line represents the induced-metric Green kernel.

Using the vertices in Eq.~(\ref{eq:fluiddiagrams}) together with the induced-metric Green kernel, the induced-metric exchange interaction between two fluid perturbations becomes
\begin{flalign}
\begin{split}
	\mathrm{Fig.~} \ref{fig:wide2} \rm{a}  
    =& \, \frac{1}{2} \int d^4x \sqrt{-g}  \int d^4x' \sqrt{-g'} \\
     & \; \; \times \Big(   \delta \rho(x) J^{\delta \rho  \delta \rho} (x,x') \delta \rho(x')\\
    & \qquad  +  \delta \rho (x) J^{\delta \rho  \xi}_{j} (x,x') w_0 (x') \xi^j(x') \\
    & \qquad   + w_0 (x)  \xi^i (x) J^{\xi  \delta \rho}_i (x,x') \delta \rho(x') \\
    & \qquad  +  w_0 (x)  \xi^i(x) J^{\xi \xi}_{ij} (x,x') w_0 (x') \xi^j (x')  \Big).
\end{split}
\label{eq:diagramcompinside}
\end{flalign}
\noindent  where we have denoted $\xi^i (x) = \alpha (x) e^{i}_I(x) \xi^I(x)$ for compactness. The kernels appearing in Eq.~(\ref{eq:diagramcompinside}) are defined by 
\begin{widetext}
\begin{flalign}
    J^{ \delta \rho  \delta \rho} (x, x') =& \frac{1}{2}  \Big( g^{00} (x) \langle h_{00}^{\rm in} (x) h_{00}^{\rm in} (x') \rangle g^{00} (x') + c_s^2(x)  g^{ij} (x) \langle h_{ij}^{\rm in} (x) h_{kl}^{\rm in} (x') \rangle g^{kl} (x') c_s^2 (x')\\
    & \quad \; - c_s^2 (x) g^{ij} (x) \langle h_{ij}^{\rm in} (x) h_{00}^{\rm in} (x') \rangle g^{00} (x') - g^{00} (x) \langle h_{00}^{\rm in} (x) h_{kl}^{\rm in} (x') \rangle g^{kl} (x') c_s^2 (x')\Big)\nonumber \\
    J^{\delta \rho \xi }_{k} (x,x') =&  \left( g^{00}(x) \langle h_{00}^{\rm in} (x) \dot{h}_{k0}^{\rm in}(x')  \rangle  g^{00} (x') - c_s^2 (x)  g^{ij} (x) \langle h_{ij}^{\rm in} \dot{h}_{k0}^{\rm in} \rangle g^{00} (x')  \right)\\
    J^{ \xi \delta \rho }_{i} (x,x') =& \left(  g^{00}(x) \langle \dot{h}_{i0}^{\rm in}(x) h_{00}^{\rm in}(x')  \rangle g^{00} (x') -    g^{00} (x)  \langle \dot{h}_{i0}^{\rm in} h_{kl}^{\rm in} \rangle g^{kl} (x') c_s^2 (x')\right)\\
    J_{ij}^{\xi \xi} (x,x') =&   g^{00}(x) \langle \dot{h}_{i0}^{\rm in}(x) \dot{h}_{j0}^{\rm in} (x') \rangle g^{00}(x') 
\end{flalign}
\end{widetext}
After integrations by parts, with the relevant spatial and temporal boundary terms vanishing as discussed in the main text, we arrive at the following explicit form of the effective fluid operator:

~

\begin{widetext}
\begin{flalign}
\begin{split}
O_{IJ}  \xi^J =& \; \alpha e^i_I\Bigg\{ \nabla_i \left( \frac{c_s^2}{ w_0} \delta \rho \right) + \frac{c_s^2 \partial_i p_0}{w_0^2}  \delta \rho  \\
& \qquad \quad  + \frac{1}{2}  \left( \nabla_i + \frac{\partial_i p_0}{w_0} \right) \left(\int dt' d^3x' \sqrt{-g (x')}  \, J^{\delta \rho \delta \rho} (x,x')  \, \delta \rho (x') \right) \\
& \qquad \quad + \left(\nabla_i + \frac{\partial_i p_0}{w_0}\right) \left(\int dt' d^3x' \sqrt{-g (x')} \, J^{\delta \rho \xi}_j (x,x') \, w_0 (x') \xi^j (x') \right) \\
& \qquad \quad - \int dt' d^3x' \sqrt{-g (x')}  \left(\, J^{\xi \delta \rho}_i (x,x') \, \delta \rho (x') + 2 \, J^{\xi \xi}_{ij} (x,x') \, w_0 (x')  \xi^j (x')\right) \Bigg\}.
\end{split}
\end{flalign}
\end{widetext}
Self-adjointness follows from integrations by parts, the vanishing of the relevant boundary terms, and the symmetry of the conservative Green kernel induced by the matched conservative boundary conditions.

\subsection*{External field metric perturbation}

To obtain the dynamics of the fluid coupled to a companion, we work in the weak-field limit, $g_{\mu \nu} = \eta_{\mu \nu} + h_{\mu  \nu}$, where post-Newtonian corrections are computed using non-relativistic general relativity~\cite{Goldberger:2004jt}. In Fourier space, the Green  function is
\begin{flalign}
    G_{\rm F} (k) =  \frac{1}{k^2+ i \epsilon}.
\end{flalign}
\noindent where $k^2 = \eta_{\mu \nu} k^{\mu} k^{\nu}$, and the perturbations are expanded in plane waves $e^{i k \cdot x}$. In the binary inspiral context, we focus on potential modes with $|k^0|\ll|\boldsymbol{k}|$, and keep only the conservative, real part of the propagator. In this limit, the propagator is instantaneous,
\begin{flalign}
\begin{split}
     \begin{gathered}
    \includegraphics[width =1.8cm]{figures/propagatorscalar.eps}
    \end{gathered} =& \langle \phi (x) \phi (x') \rangle
    \\=&   4 \pi G \delta(t-t')  \int \frac{d^3 k}{(2\pi)^3} \frac{e^{i \boldsymbol{k}\cdot (\boldsymbol{x}-\boldsymbol{x}')}}{\boldsymbol{k}^2} ~, \label{eq:propN} 
\end{split}
\end{flalign}
Here the $i \epsilon$ prescription has been dropped in the conservative sector, and time-derivative corrections are treated as higher-order post-Newtonian effects~\cite{Goldberger:2004jt}.

The companion worldline coupling in the weak-field limit is 
\begin{align}
    \begin{split}
        \begin{gathered}
   \includegraphics[width = 1.5cm]{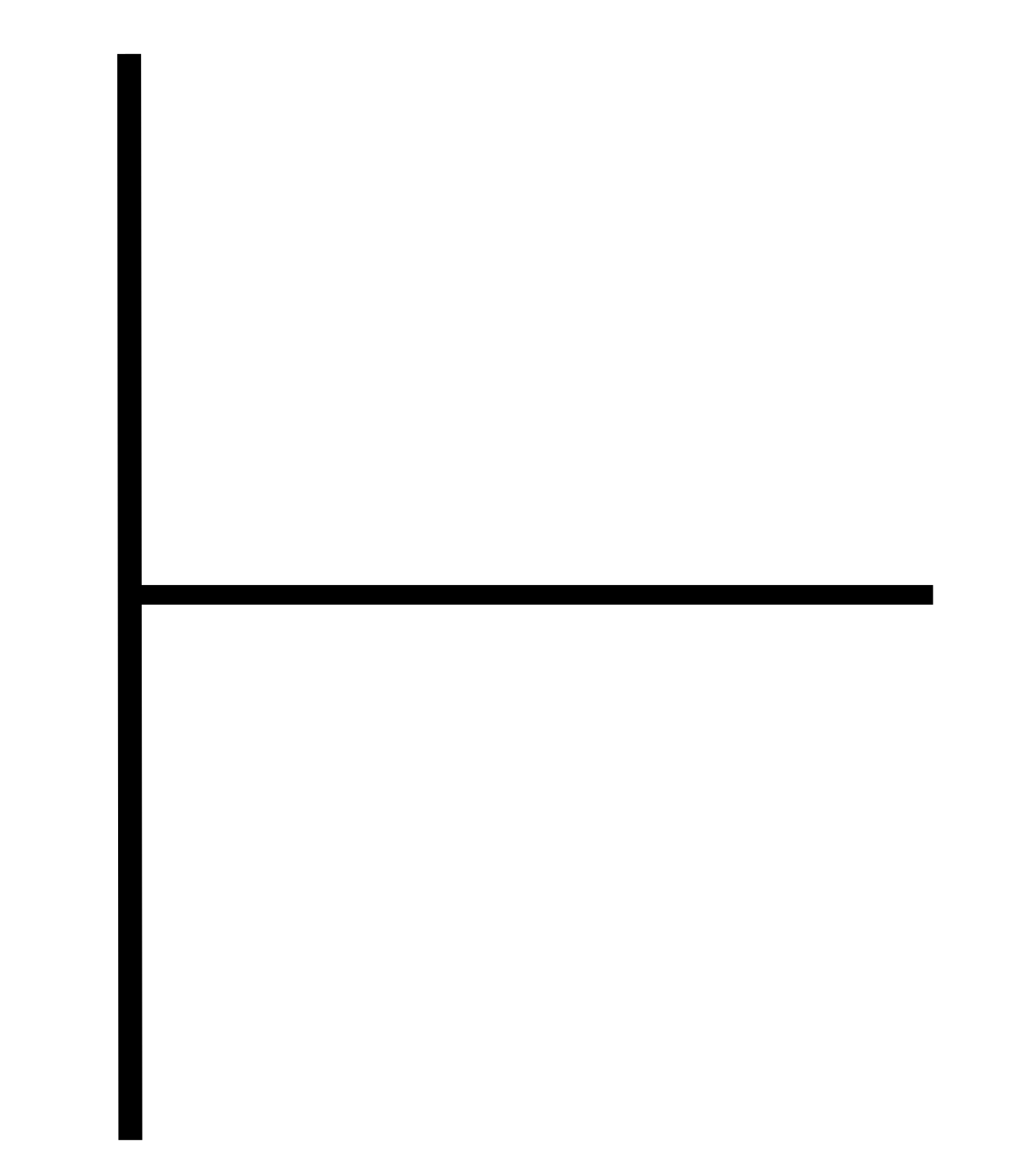}
        \end{gathered}  =&  M \int d t  \left(  \phi + O(v^2) \right),
    \end{split}
\label{eq:pointmassv}
\end{align}
\noindent where $M$ is the companion mass, $\phi = - \tfrac{1}{2} h_{00}$ is the weak-field potential mode and $O(v^2)$ denotes post-Newtonian corrections.

Integrating out the potential mode in the weak-field approximation gives
\begin{flalign}
\begin{split}
\mathrm{Fig.~}\ref{fig:wide2}\rm{b}  &=  -\int d t \int d t'   d^3 x \delta \rho (\vec{x}) \left< \phi (x) \phi (x')\right>  M  \\
    &=  \int d t d^3x  \delta \rho(\vec{x}) \Phi_{\rm ext},
\end{split}
\label{eq:diagramcomp}
\end{flalign}
\noindent with $\Phi_{\rm ext} = -G M/r$ as expected.

\section{Approximations and limitations}
\label{app:approx}

The effective description developed here is valid under a set of assumptions and breaks down when those assumptions cease to hold. We model the neutron star as a relativistic perfect fluid and restrict to the conservative linear response about a static spherical equilibrium configuration. In this regime, the pole part of the tidal response is represented by the star's discrete normal modes, so the corresponding worldline response can be organized as a set of driven harmonic oscillators. At low frequencies, the dominant contribution is typically provided by the fundamental $f$-mode, while higher modes give corrections controlled by their frequencies, normalizations, and overlap integrals. In practice, the mode sum must be truncated, and the accuracy of a finite-mode truncation deteriorates as the  frequency $\omega$ approaches the first omitted resonance. 

In the slow inspiral regime, metric perturbations can be regarded as approximately static,
\(\dot{\phi}_{\rm ext}\simeq 0\). Dynamical tidal effects become relevant as the orbital frequency approaches the star's eigenfrequencies. For a binary in a circular orbit, the orbital frequency is given at leading order by
$\Omega = \sqrt{ G (M_1 +M_2) / r^3}$. 
Resonance is reached when the relevant driving frequency approaches the star's eigenfrequencies, in particular the fundamental-mode frequency \(\omega_f\). Near resonance, the mode amplitude can grow, and nonlinearities may become important, signaling the breakdown of the linear response model.

The relevance of nonlinearities can be estimated parametrically by expanding the fluid perturbation action to $O(\pi^3)$. An illustrative cubic contribution is
\begin{flalign}
    S \sim \int d^4 x \sqrt{-g} \frac{1}{ w_0^2} c_s^2  (\delta \rho)^3. 
\end{flalign}
\noindent Using the fluid perturbation eigenfunctions, the time-dependent amplitude can be rewritten in terms of the corresponding quadrupole variable $Q$. For a single mode, this gives schematically
\begin{flalign}
    S \sim \int dt \frac{1}{(M_{\star} R_{\star}^2 \mathcal{N}_{n} \lambda_n)^{3/2} \omega_n^3}Q^3 \int dr r^2 e^{\Phi+\Lambda} \frac{c_s^2}{w_0^2}(\delta \rho_{n})^3  
\end{flalign}

Expressing the radial integral as a dimensionless cubic overlap,
\begin{flalign}
    \mathcal{J}_{n} = \frac{1}{M_{\star}} \int dr r^2 e^{\Phi+\Lambda} \frac{c_s^2}{w_0^2} (\delta \rho_n)^3,   
\end{flalign}
\noindent and canonically normalizing the quadrupolar variable, the induced cubic worldline interaction scales parametrically as
\begin{flalign}
    L_{Q^3} \sim \frac{1}{ R_{\star}^3 M_{\star}^{1/2} } \frac{  \mathcal{J}_{n}}{\mathcal{N}_{n}^{3/2}} Q^3. 
\end{flalign}
This estimate gives only the parametric size of the leading nonlinear correction. A precise determination of the nonlinear breakdown point for close binaries would require evolving the driven mode amplitudes \(q_n(t)\), or equivalently the quadrupolar variables \(Q_n(t)\), along a binary inspiral. Such an analysis is beyond the scope of the present work.

% Produces the bibliography via BibTeX.

\bibliography{apssamp}

@PREAMBLE{
 "\providecommand{\noopsort}[1]{}" 
 # "\providecommand{\singleletter}[1]{#1}%" 
}

@article{Dubovsky:2011sj,
    author = "Dubovsky, Sergei and Hui, Lam and Nicolis, Alberto and Son, Dam Thanh",
    title = "{Effective field theory for hydrodynamics: thermodynamics, and the derivative expansion}",
    eprint = "1107.0731",
    archivePrefix = "arXiv",
    primaryClass = "hep-th",
    doi = "10.1103/PhysRevD.85.085029",
    journal = "Phys. Rev. D",
    volume = "85",
    pages = "085029",
    year = "2012"
}

@article{HegadeKR:2024agt,
    author = "Hegade K. R., Abhishek and Ripley, Justin L. and Yunes, Nicol{\'a}s",
    title = "{Dynamical tidal response of nonrotating relativistic stars}",
    eprint = "2403.03254",
    archivePrefix = "arXiv",
    primaryClass = "gr-qc",
    doi = "10.1103/PhysRevD.109.104064",
    journal = "Phys. Rev. D",
    volume = "109",
    number = "10",
    pages = "104064",
    year = "2024"
}

@article{HegadeKR:2025qwj,
    author = "Hegade K. R., Abhishek and Kwon, K. J. and Venumadhav, Tejaswi and Yu, Hang and Yunes, Nicol{\'a}s",
    title = "{Relativistic and Dynamical Love Numbers}",
    eprint = "2507.10693",
    archivePrefix = "arXiv",
    primaryClass = "gr-qc",
    doi = "10.1103/1wdp-6x27",
    journal = "Phys. Rev. Lett.",
    volume = "136",
    number = "7",
    pages = "071401",
    year = "2026"
}

@article{Chakraborty:2026qru,
    author = "Chakraborty, Sumanta and Pani, Paolo",
    title = "{Tidal Response of Compact Objects}",
    eprint = "2604.08679",
    journal = "arXiv preprint",
    archivePrefix = "arXiv",
    primaryClass = "gr-qc",
    month = "4",
    year = "2026"
}

@article{Zhou:2025lzg,
    author = "Zhou, Zihan and Tomaselli, Giovanni Maria and Mart{\'\i}nez-Rodr{\'\i}guez, Irvin and Li, Jingping",
    title = "{Modeling Tidal Disruptions with Dynamical Tides}",
    eprint = "2504.16025",
    archivePrefix = "arXiv",
    primaryClass = "astro-ph.HE",
    doi = "10.3847/1538-4357/adf190",
    journal = "Astrophys. J.",
    volume = "990",
    number = "2",
    pages = "198",
    year = "2025"
}

@article{Andersson:2025iyd,
    author = "Andersson, Nils and Counsell, Rhys and Gittins, Fabian and Ghosh, Suprovo",
    title = "{Tidal response of a relativistic star}",
    eprint = "2511.05139",
    archivePrefix = "arXiv",
    primaryClass = "gr-qc",
    doi = "10.1103/x4yr-x1c4",
    journal = "Phys. Rev. D",
    volume = "113",
    number = "6",
    pages = "064051",
    year = "2026"
}

@article{Apostolidis:2026qsg,
    author = "Apostolidis, Thomas and De Luca, Valerio and Gualtieri, Leonardo and Katagiri, Takuya and Pani, Paolo and Santoni, Luca",
    title = "{Dynamical Tidal Response of Neutron Stars: from Effective Field Theory to Gravitational Waveforms}",
    journal = "arXiv preprint",
    eprint = "2606.19446",
    archivePrefix = "arXiv",
    primaryClass = "gr-qc",
    month = "6",
    year = "2026"
}

@article{Saketh:2026trm,
    author = "Saketh, M. V. S. and Ghosh, Suprovo and Andersson, Nils",
    title = "{Dynamical tidal response of neutron stars via scattering amplitudes}",
    journal = "arXiv preprint",
    eprint = "2606.14405",
    archivePrefix = "arXiv",
    primaryClass = "gr-qc",
    month = "6",
    year = "2026"
}

@article{HegadeKR:2026kku,
    author = "Hegade K. R., Abhishek and Kwon, K. J. and Venumadhav, Tejaswi and Yu, Hang and Yunes, Nicolas",
    title = "{The Good, the Bad, and the Subtle: Relativistic mode sums for neutron-star tidal response}",
    eprint = "2605.08569",
    journal = "arXiv preprint",
    archivePrefix = "arXiv",
    primaryClass = "gr-qc",
    month = "5",
    year = "2026"
}

@article{Detweiler:1985zz,
    author = "Detweiler, Steven L. and Lindblom, L.",
    title = "{On the nonradial pulsations of general relativistic stellar models}",
    doi = "10.1086/163127",
    journal = "Astrophys. J.",
    volume = "292",
    pages = "12--15",
    year = "1985"
}

@article{Rodriguez:2026iot,
    author = "Rodr{\'\i}guez, Mar{\'\i}a J. and Santoni, Luca and Solomon, Adam R.",
    title = "{Love numbers of black holes and compact objects}",
    journal = "arXiv preprint",
    eprint = "2604.08653",
    archivePrefix = "arXiv",
    primaryClass = "gr-qc",
    month = "4",
    year = "2026"
}

@article{Pnigouras:2022zpx,
    author = "Pnigouras, Pantelis and Gittins, Fabian and Nanda, Amlan and Andersson, Nils and Jones, David Ian",
    title = "{Rotating Love: The dynamical tides of spinning Newtonian stars}",
    eprint = "2205.07577",
    archivePrefix = "arXiv",
    primaryClass = "gr-qc",
    doi = "10.1093/mnras/stad3593",
    journal = "Mon. Not. Roy. Astron. Soc.",
    volume = "527",
    pages = "8409--8428",
    year = "2024"
}

@article{Lai:1993di,
    author = "Lai, Dong",
    title = "{Resonant oscillations and tidal heating in coalescing binary neutron stars}",
    eprint = "astro-ph/9404062",
    archivePrefix = "arXiv",
    reportNumber = "CRSR-1064",
    doi = "10.1093/mnras/270.3.611",
    journal = "Mon. Not. Roy. Astron. Soc.",
    volume = "270",
    pages = "611",
    year = "1994"
}

@article{Lindblom:1983ps,
    author = "Lindblom, L and Detweiler, Steven L.",
    title = "{The quadrupole oscillations of neutron stars}",
    doi = "10.1086/190884",
    journal = "Astrophys. J. Suppl.",
    volume = "53",
    pages = "73--92",
    year = "1983"
}

@article{Dubovsky:2005xd,
    author = "Dubovsky, S. and Gregoire, T. and Nicolis, A. and Rattazzi, R.",
    title = "{Null energy condition and superluminal propagation}",
    eprint = "hep-th/0512260",
    archivePrefix = "arXiv",
    reportNumber = "CERN-PH-TH-2005-265, HUTP-05-A0056",
    doi = "10.1088/1126-6708/2006/03/025",
    journal = "JHEP",
    volume = "03",
    pages = "025",
    year = "2006"
}

@article{Hinderer:2007mb,
    author = "Hinderer, Tanja",
    title = "{Tidal Love numbers of neutron stars}",
    eprint = "0711.2420",
    archivePrefix = "arXiv",
    primaryClass = "astro-ph",
    doi = "10.1086/533487",
    journal = "Astrophys. J.",
    volume = "677",
    pages = "1216--1220",
    year = "2008"
}

@article{Binnington:2009bb,
    author = "Binnington, Taylor and Poisson, Eric",
    title = "{Relativistic theory of tidal Love numbers}",
    eprint = "0906.1366",
    archivePrefix = "arXiv",
    primaryClass = "gr-qc",
    doi = "10.1103/PhysRevD.80.084018",
    journal = "Phys. Rev. D",
    volume = "80",
    pages = "084018",
    year = "2009"
}

@article{Nicolis:2013lma,
    author = "Nicolis, Alberto and Penco, Riccardo and Rosen, Rachel A.",
    title = "{Relativistic Fluids, Superfluids, Solids and Supersolids from a Coset Construction}",
    eprint = "1307.0517",
    archivePrefix = "arXiv",
    primaryClass = "hep-th",
    doi = "10.1103/PhysRevD.89.045002",
    journal = "Phys. Rev. D",
    volume = "89",
    number = "4",
    pages = "045002",
    year = "2014"
}

@article{Chakrabarti:2013xza,
    author = "Chakrabarti, Sayan and Delsate, Térence and Steinhoff, Jan",
    archivePrefix = "arXiv",
    doi = "10.1103/PhysRevD.88.084038",
    eprint = "1306.5820",
    journal = "Phys.\ Rev.\ D",
    pages = "084038",
    primaryClass = "gr-qc",
    title = "{Effective action and linear response of compact objects in Newtonian gravity}",
    volume = "88",
    year = "2013"
}

@article{Poisson:2023xsr,
    author = "Pitre, Tristan and Poisson, Eric",
    title = "{General relativistic dynamical tides in binary inspirals without modes}",
    eprint = "2311.04075",
    archivePrefix = "arXiv",
    primaryClass = "gr-qc",
    doi = "10.1103/PhysRevD.109.064004",
    journal = "Phys. Rev. D",
    volume = "109",
    number = "6",
    pages = "064004",
    year = "2024"
}

@article{Oppenheimer:1939ne,
    author = "Oppenheimer, J. R. and Volkoff, G. M.",
    title = "{On massive neutron cores}",
    doi = "10.1103/PhysRev.55.374",
    journal = "Phys. Rev.",
    volume = "55",
    pages = "374--381",
    year = "1939"
}

@article{Tolman:1939jz,
    author = "Tolman, Richard C.",
    title = "{Static solutions of Einstein's field equations for spheres of fluid}",
    doi = "10.1103/PhysRev.55.364",
    journal = "Phys. Rev.",
    volume = "55",
    pages = "364--373",
    year = "1939"
}

@article{Lindblom:1997un,
    author = "Lindblom, Lee and Mendell, Gregory and Ipser, James R.",
    title = "{Relativistic stellar pulsations with near - zone boundary conditions}",
    eprint = "gr-qc/9704046",
    archivePrefix = "arXiv",
    doi = "10.1103/PhysRevD.56.2118",
    journal = "Phys. Rev. D",
    volume = "56",
    pages = "2118--2126",
    year = "1997"
}

@article{Damour:2009vw,
    author = "Damour, Thibault and Nagar, Alessandro",
    title = "{Relativistic tidal properties of neutron stars}",
    eprint = "0906.0096",
    archivePrefix = "arXiv",
    primaryClass = "gr-qc",
    doi = "10.1103/PhysRevD.80.084035",
    journal = "Phys. Rev. D",
    volume = "80",
    pages = "084035",
    year = "2009"
}

@article{Flanagan:2007ix,
    author = "Flanagan, Eanna E. and Hinderer, Tanja",
    title = "{Constraining neutron star tidal Love numbers with gravitational wave detectors}",
    eprint = "0709.1915",
    archivePrefix = "arXiv",
    primaryClass = "astro-ph",
    doi = "10.1103/PhysRevD.77.021502",
    journal = "Phys. Rev. D",
    volume = "77",
    pages = "021502",
    year = "2008"
}

@article{Goldberger:2004jt,
    author = "Goldberger, Walter D. and Rothstein, Ira Z.",
    archivePrefix = "arXiv",
    doi = "10.1103/PhysRevD.73.104029",
    eprint = "hep-th/0409156",
    journal = "Phys.\ Rev.\ D",
    pages = "104029",
    reportNumber = "UCSD-PTH-04-17, CMU-HEP-04-06",
    title = "{An Effective field theory of gravity for extended objects}",
    volume = "73",
    year = "2006"
}

@article{Goldberger:2005cd,
    author = "Goldberger, Walter D. and Rothstein, Ira Z.",
    title = "{Dissipative effects in the worldline approach to black hole dynamics}",
    eprint = "hep-th/0511133",
    archivePrefix = "arXiv",
    doi = "10.1103/PhysRevD.73.104030",
    journal = "Phys. Rev. D",
    volume = "73",
    pages = "104030",
    year = "2006"
}

@article{Delacretaz:2014oxa,
    author = "Delacrétaz, Luca V. and Endlich, Solomon and Monin, Alexander and Penco, Riccardo and Riva, Francesco",
    archivePrefix = "arXiv",
    doi = "10.1007/JHEP11(2014)008",
    eprint = "1405.7384",
    journal = "JHEP",
    pages = "008",
    primaryClass = "hep-th",
    title = "{(Re-)Inventing the Relativistic Wheel: Gravity, Cosets, and Spinning Objects}",
    volume = "11",
    year = "2014"
}

\end{document}